\documentclass{jpsj-suppl}
\usepackage{times}
\usepackage{color}
\usepackage{ulem}
\usepackage{bm}

\title{Invariant Energy Levels and Flat Band Engineering in a Kondo Lattice Model on Geometrically Frustrated Lattices}

\author{Masafumi Udagawa\thanks{E-mail address: udagawa@ap.t.u-tokyo.ac.jp}, Hiroaki Ishizuka, and Yukitoshi Motome 
}
\inst{Department of Applied Physics, University of Tokyo, 7-3-1, Hongo, Bunkyo-ku, Tokyo 113-8656 
}
\abst{We show the existence of invariant energy levels 
in a Kondo lattice model 
on an isolated complete graph, such as a triangle and a tetrahedron. 
These energy levels always have fixed eigenenergies $t \pm\frac{J}{2}$, irrespective of the configuration of localized moments ($t$ is the transfer integral of conduction electrons and $J$ is the spin-charge coupling constant). 
We also extend the analysis to geometrically frustrated lattices by using the complete graphs as basic building blocks. 
We show that the construction rule 
for the invariant energy levels
leads to the existence condition of localized states, if the model is defined on the triangle-based line graphs, such as a kagome lattice. 
We further propose a procedure 
of engineering isolated flat bands 
with broken time-reversal symmetry, which are separated from other dispersive bands with finite energy gaps. 
}

\kword{Geometrical frustration, flat band, Kondo lattice model}

\begin{document}
\maketitle

\section{Introduction}
Conduction electrons sometimes exhibit anomalous properties on geometrically frustrated lattices. Geometrical frustration causes characteristic interference in electron wave functions, 
and alters the nature of Bloch states considerably. A representative example is a flat band. 
An energy band 
with completely flat dispersion appears in a non-interacting tight-binding model defined on a class of geometrically frustrated lattices
composed of complete graphs, such as triangles and tetrahedra. A number of rigorous results associated with such flat bands have 
been obtained\cite{Mielke91_1, Mielke91_2, Mielke92, Brandt92, Katsura10, Bergman08}.

The existence of a flat band immediately means 
the divergence of the density of states, implying an extreme sensitivity to electron interactions. In the Hubbard model, 
it was rigorously shown that the onsite Coulomb repulsion induces the
ferromagnetic ground state\cite{Mielke91_1, Mielke91_2} and spin liquid ground state\cite{Brandt92, Mielke92}. 
Among the consequences of interaction effects, recently, possible emergence of a fractional Chern insulator 
was pointed out\cite{Tang11,Sun11,Neupert11}. Provided that a sufficiently flat band has 
a nonzero Chern number, 
a repulsive interaction drives the system into a fractional quantum Hall state. This fascinating proposal, however, still remains at a theoretical level, partly because a recipe for
such flat bands has not been firmly established.

In this 
contribution, we consider the classical Kondo lattice model as a possible stage to realize flat bands with broken time-reversal symmetry. 
The classical Kondo lattice model describes the interaction between conduction electrons and localized classical magnetic moments.
This model has been intensively studied as a representative model for perovskite manganese oxides\cite{Tokura00}. 
Recently, it has also been studied on geometrically frustrated lattices, partly motivated by the interesting properties in 
pyrochlore Iridates\cite{Udagawa12, Udagawa13} and Molybdates\cite{Motome}. 
In 
the present study, 
we first show the existence of invariant energy levels appearing in the Kondo lattice model on an isolated 
complete graph,
which have fixed eigenenergies 
irrespective of the configuration of localized moments.
Then, 
extending our analysis to lattice systems, 
we establish 
a procedure of 
engineering a flat band with broken time-reversal symmetry, by making use of the 
construction rule of the invariant energy levels on an isolated cluster.

\section{Model}
\label{Model}
We consider 
a ferromagnetic Kondo lattice model, whose Hamiltonian is written as
\begin{align}
\mathcal{H} = -t\sum\limits_{\langle i,j\rangle, \alpha} (c_{i\alpha}^{\dag}c_{j\alpha} + {\rm H.c.}) - &J\sum\limits_{i,\alpha,\beta} c_{i\alpha}^{\dag}\bigl(\frac{1}{2}{\bm\sigma}_{\alpha\beta}\bigr)c_{i\beta}\cdot{\mathbf S}_i.
\label{Ham}
\end{align}
Here, the first term describes the kinetic energy of electrons (spin-$1/2$ fermions). The sum $\langle i,j\rangle$ is taken over the
nearest-neighbor 
sites on a lattice or graph. The second term represents the coupling between conduction
electron spins ($\bm{\sigma}_{\alpha\beta}$ is the vector of Pauli matrices) and localized magnetic moments ${\mathbf S}_i$. 
In this model, we assume ${\mathbf S}_i$ to be 
classical spins with a fixed amplitude of $
|{\mathbf S}_i|=1
$.
Hereafter, we take $t = 1$ as an energy unit, and adopt the convention: $J>0$. (The sign of $J$ can be switched by time-reversal operation.)

\begin{figure}[h]
\begin{center}
\includegraphics[width=0.8\textwidth]{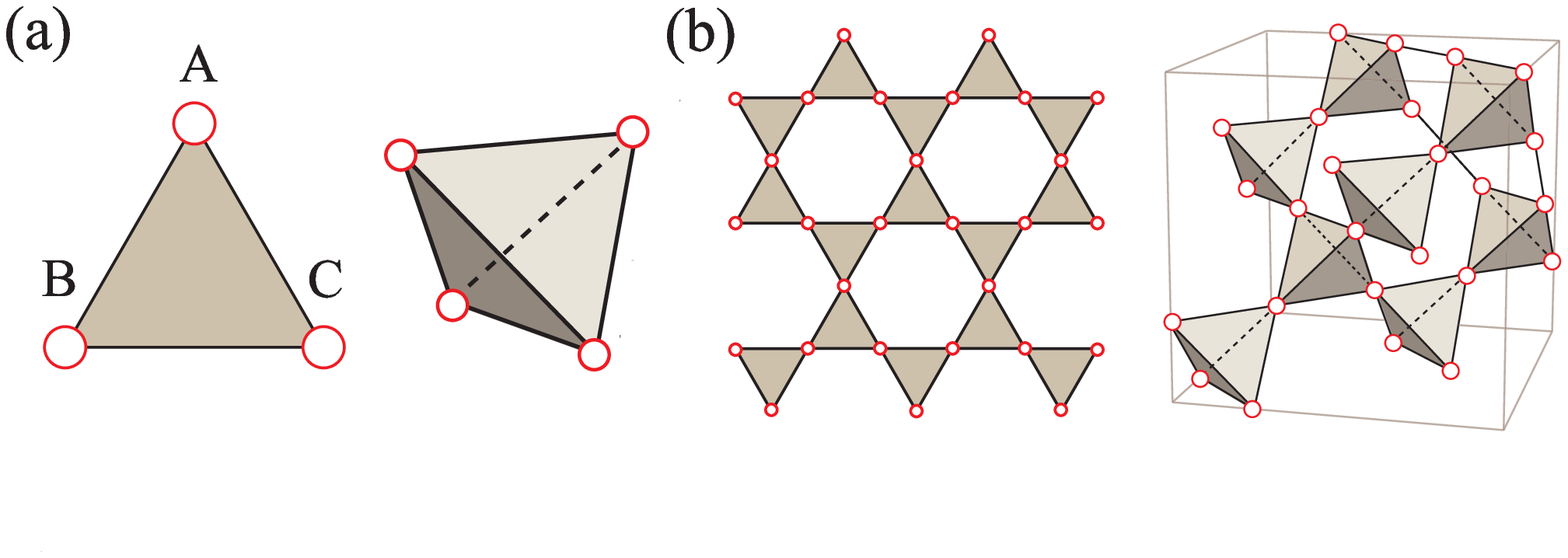}
\end{center}
\caption{\label{graph} 
(Color online) (a) 
Triangle and 
tetrahedron as examples of complete graphs, and (b) kagome and pyrochlore lattices, made by connecting triangles and tetrahedra as building blocks, respectively.
}
\end{figure}

We consider the Kondo lattice model on (i) isolated complete graphs and (ii) 
lattices composed of complete graphs, as 
exemplified in Fig.~\ref{graph}. 
The complete graph is defined as a graph composed of $n_c$ sites, where each site is connected to the other $n_c-1$ sites.
A triangle ($n_c=3$) and a tetrahedron ($n_c=4$) give the simplest examples [Fig.~\ref{graph}(a)].
Typical geometrically frustrated lattices are constructed by connecting the complete graphs 
with sharing their corners. For example, the kagome lattice is composed of triangles and pyrochlore lattice is made of tetrahedra [Fig.~\ref{graph}(b)]. 
This class of lattices is known for the localized states appearing in the non-interacting tight-binding models defined on them. 

\section{Results}
In this section, we present our main results.
In 
Sec.~\ref{isolated_graph_Jinf} and Sec.~\ref{isolated_graph_Jf}, we show the existence of ``invariant energy levels" in 
the Kondo lattice model on an isolated complete graph for $J=\infty$ and finite $J$, respectively. In 
Sec.~\ref{kagome}, we show that the construction rule 
for these levels leads to the existence condition of localized modes on the kagome lattice. We further 
establish the way of engineering flat bands with broken time-reversal symmetry, by making use of the 
construction rule of the invariant energy levels.

\subsection{Isolated complete graph: $J=\infty$}
\label{isolated_graph_Jinf}
First, 
we consider the system with $J=\infty$. 
In this strong coupling limit, the complete spectrum has 
been obtained by Penc and Lacaze\cite{Penc99}, including the case of quantum localized spin. Nevertheless, we 
provide a detailed derivation here, since this limit gives an illustrative example which is easily generalized to finite $J$,
as we will show in the next section. As a specific example, we consider a single triangle [Fig.~\ref{graph}(a)] on which the Hamiltonian (\ref{Ham}) is defined.
In the limit of $J 
\to \infty$, the one-particle eigenenergies 
of the Hamiltonian (\ref{Ham}) are split into two sectors with an energy gap of the magnitude $\sim J$,
where the lower-(higher-)energy sector is composed of the electronic states whose spins are parallel (antiparallel) to the localized moments at each site.
Accordingly, the effective Hamiltonian for the lower-energy sector can be written as
\begin{eqnarray}
\mathcal{H}_{\rm eff} = -t\sum\limits_{\langle i,j\rangle, \alpha}(\langle\chi_+({\mathbf S}_i)|\chi_+({\mathbf S}_j)\rangle a^{\dag}_{i+}a_{j+} + \rm{H.c.}),
\label{Heff}
\end{eqnarray}
up to the order of $t/J$\cite{Anderson55}. 
Here we have introduced the spinor eigenstates $|\chi_{\pm}({\mathbf S}_i)\rangle\equiv\chi_{\pm\uparrow}({\mathbf S}_i)|\uparrow\rangle + \chi_{\pm\downarrow}({\mathbf S}_i)|\downarrow\rangle$, which satisfy
${\bm\sigma}\cdot{\mathbf S}_i|\chi_{\pm}({\mathbf S}_i)\rangle = \pm|\chi_\pm({\mathbf S}_i)\rangle$, 
and are normalized so that $\langle\chi_{s}({\mathbf S}_i)|\chi_{s'}({\mathbf S}_i)\rangle = \delta_{s,s'}$ ($s, s'=\pm$).
The operator to annihilate 
an electron in the 
state $|\chi_{\pm}({\mathbf S}_i)\rangle$ is defined as $a_{i\pm}\equiv\sum_{\alpha=\uparrow,\downarrow}\chi^*_{\pm\alpha}({\mathbf S}_i)c_{i\alpha}$.

On a single 
triangle, each site is connected with the other 
two sites. Accordingly, the effective Hamiltonian (\ref{Heff}) can be explicitly written in the matrix form, as
\begin{eqnarray}
\mathcal{H}_{\rm eff} = -t
\vec{a}_+^{\dag}\hat{\mathcal{K}}^{++}\vec{a}_+ + t\vec{a}_+^{\dag}\vec{a}_+.
\label{Heff2}
\end{eqnarray}
Here, we introduced the $3\times3$ matrix $\hat{\mathcal{K}}^{ss'}$ ($s, s'=\pm$), whose elements are given by $[\hat{\mathcal{K}}^{ss'}]_{jj'} = \langle\chi_s({\mathbf S}_j)|\chi_{s'}({\mathbf S}_{j'})\rangle$, and a vector of creation operators, $\vec{a}_{\pm}^{\dag} = [a_{{\rm A}\pm}^{\dag}, a_{{\rm B}\pm}^{\dag}, a_{{\rm C}\pm}^{\dag}]$, where ${\rm A}$, ${\rm B}$, and ${\rm C}$ correspond to the site indices in Fig.~\ref{graph}(a).
Note that the second term in Eq.~(\ref{Heff2}) cancels the diagonal terms in the first term. A crucial observation is that $\hat{\mathcal{K}}^{++}$ can be written
in a factorized form, 
\begin{eqnarray}
\hat{\mathcal{K}}^{++} = \hat{\chi}_+^{\dag}\hat{\chi}_+; \hspace{0.5cm} \hat{\chi}_+ = 
\begin{bmatrix}
\chi_{+\uparrow}({\mathbf S}_{\rm A}) & \chi_{+\uparrow}({\mathbf S}_{\rm B}) & \chi_{+\uparrow}({\mathbf S}_{\rm C})\\
\chi_{+\downarrow}({\mathbf S}_{\rm A}) & \chi_{+\downarrow}({\mathbf S}_{\rm B}) & \chi_{+\downarrow}({\mathbf S}_{\rm C})\\
\end{bmatrix}.
\label{eq:K++}
\end{eqnarray}
Hence, if a non-vanishing one-particle state $|\Psi\rangle\equiv(\Psi_{\rm A}a_{{\rm A}+}^{\dag} + \Psi_{\rm B}a_{{\rm B}+}^{\dag} + \Psi_{\rm C}a_{{\rm C}+}^{\dag})|0\rangle$ ($|0\rangle$: the vacuum state) satisfies the following relation, 
\begin{eqnarray}
\Psi_{\rm A}\chi_{+\alpha}({\mathbf S}_{\rm A}) + \Psi_{\rm B}\chi_{+\alpha}({\mathbf S}_{\rm B}) + \Psi_{\rm C}\chi_{+\alpha}({\mathbf S}_{\rm C}) = 0 
\quad {\rm for} \ \ \alpha=\uparrow {\rm and} \downarrow,
\label{constraint}
\end{eqnarray}
$\hat{\mathcal{K}}^{++}$ annihilates this state $|\Psi\rangle$. 
We, therefore, obtain
$\mathcal{H}_{\rm eff}|\Psi\rangle = t|\Psi\rangle$; that is,
$|\Psi\rangle$ is 
an eigenstate of $\mathcal{H}_{\rm eff}$ with 
eigenvalue $t$, {\it irrespective of the spin configurations $({\mathbf S}_{\rm A}, {\mathbf S}_{\rm B}, {\mathbf S}_{\rm C})$}.
Since the matrix 
$\hat{\mathcal{K}}^{++} $ is semi-positive definite, this state gives the highest-energy eigenstate of $\mathcal{H}_{\rm eff}$.  

Here, we remark on the degeneracy of states. 
Equation~(\ref{constraint}) imposes two independent constraints on the three coefficients $\Psi_j$. 
Consequently, 
there remains only one degree of freedom, i.e., the state $|\Psi \rangle$ has no degeneracy in this case. It is straightforward to generalize the argument to the Kondo lattice model with spin-$S$ fermions defined on a complete graph composed of $n_c$ sites. In this general case, the coefficient $\Psi_j$ has $n_c$ components, while the constraint in Eq.~(\ref{constraint}) is composed of $2S+1$ independent equations,
resulting in 
$n_c-(2S+1)$-fold 
degeneracy 
in the eigenstate with eigenenergy $t$.

\subsection{Isolated complete graph: finite $J$}
\label{isolated_graph_Jf}
\begin{figure}[h]
\begin{center}
\includegraphics[width=0.99\textwidth]{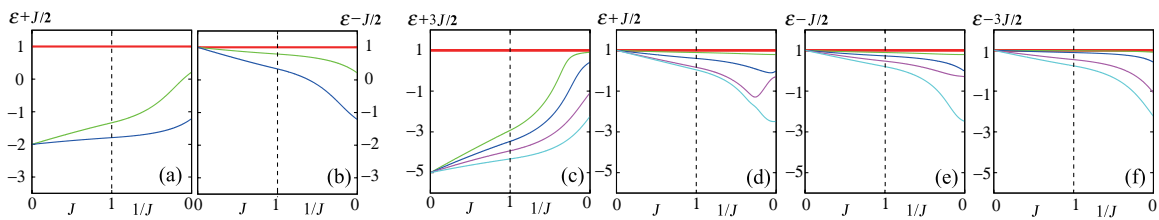}
\end{center}
\caption{\label{TetrahedronLevel} 
(Color online) 
(a), (b) $J$ dependence of energy levels 
of the Kondo lattice model on a single triangle, corresponding to lower and upper three levels, respectively. The spin configuration is assumed to be ${\mathbf S}_i=(\sin\theta_i\cos\phi_i, \sin\theta_i\sin\phi_i, \cos\theta_i)$ 
with $(\theta_{\rm A}, \phi_{\rm A})=(3\pi/5, \pi/10)$, $(\theta_{\rm B}, \phi_{\rm B})=(9\pi/10, 6\pi/5)$ and $(\theta_{\rm C}, \phi_{\rm C})=(2\pi/5, 3\pi/2)$. The top levels are located at $\varepsilon=-\frac{J}{2}+t$ and $\frac{J}{2}+t$, respectively. (c)-(f) depict the energy levels of Kondo lattice model with spin-3/2  
fermions on a complete graph composed of $n_c=6$ sites. The spin configuration is 
chosen as $(\theta_1, \phi_1)=(\pi/4, \pi/3)$, $(\theta_2, \phi_2)=(\pi/6, 5\pi/3)$, $(\theta_3, \phi_3)=(2\pi/7, 3\pi/8)$, $(\theta_4, \phi_4)=(2\pi/5, 4\pi/3)$, $(\theta_5, \phi_5)=(3\pi/7, \pi/2)$ and $(\theta_6, \phi_6)=(\pi/2, 8\pi/5)$. (c), (d), (e), and (f) correspond to the 1st to 6th, 7th to 12th, 13th to 18th, and 19th to 24th energy levels, respectively. 
The top levels are located at $\varepsilon=S_zJ + t$ with $S_z=-3/2$, $-1/2$, $1/2$, and $3/2$, each with 
twofold degeneracy.}
\end{figure}

Next, we consider the case of finite $J$. In this case, the parts of Hilbert space spanned by $|\chi_{+}({\mathbf S}_i)\rangle$ and $|\chi_{-}({\mathbf S}_i)\rangle$ are mixed with each other. Consequently, taking 
$|\chi_{\pm}({\mathbf S}_i)\rangle$ as local spinor bases, the Hamiltonian (\ref{Ham}) can be written as
\begin{eqnarray}
 \mathcal{H} &= -t
\begin{bmatrix}
\vec{a}_{+}^{\dag} & \vec{a}_{-}^{\dag}
\end{bmatrix}
(\hat{\mathcal{K}}_{\rm T} + \hat{\mathcal{K}}_{\rm J})
\begin{bmatrix}
\vec{a}_{+}\\ 
\vec{a}_{-}
\end{bmatrix}
+t(\vec{a}_{+}^{\dag}\vec{a}_{+} + \vec{a}_{-}^{\dag}\vec{a}_{-}). 
\label{Heff4}
\end{eqnarray}
Here, $\hat{\mathcal{K}}_{\rm T}$ and $\hat{\mathcal{K}}_{\rm J}$ are $6 \times 6$ matrices defined by
\begin{eqnarray}
\hat{\mathcal{K}}_{\rm T} = 
\begin{bmatrix}
\hat{\mathcal{K}}^{++} & \hat{\mathcal{K}}^{+-}\\
\hat{\mathcal{K}}^{-+} & \hat{\mathcal{K}}^{--}
\end{bmatrix}=
\begin{bmatrix}
\hat{\chi}_+^{\dag}\\
\hat{\chi}_-^{\dag}
\end{bmatrix}
\begin{bmatrix}
\hat{\chi}_+ & \hat{\chi}_-
\end{bmatrix},
\hspace{1cm}
\hat{\mathcal{K}}_{\rm J} = \frac{J}{2}
\begin{bmatrix}
\hat{1} & \hat{0}\\
\hat{0} & -\hat{1}
\end{bmatrix},
\end{eqnarray}
where $\hat{\chi}_-$ is defined similarly to $\hat{\chi}_+$ in Eq.~(\ref{eq:K++}); 
$\hat{1}$ and $\hat{0}$ are the 
unit and zero matrices of the size 
$3\times 3$. Again, due to the factorized form of $\hat{\mathcal{K}}_{\rm T}$, if a non-vanishing one-particle state $|\Psi\rangle\equiv\sum_{j={\rm A},{\rm B},{\rm C}}(\Psi_{j+}a_{j+}^{\dag} + \Psi_{j-}a_{j-}^{\dag})|0\rangle$ satisfies the conditions: 
\begin{eqnarray}
\sum_{j={\rm A},{\rm B},{\rm C}}(\chi_{+\alpha}({\mathbf S}_j)\Psi_{j+} + \chi_{-\alpha}({\mathbf S}_j)\Psi_{j-}) = 0
\quad {\rm for} \ \  \alpha=\uparrow {\rm and} \downarrow, 
\label{finiteJcondition}
\end{eqnarray}
$\hat{\mathcal{K}}_{\rm T}$ annihilates the state $|\Psi\rangle$. For the state $|\Psi\rangle$ to be an eigenstate of $\mathcal{H}$, $|\Psi\rangle$ also has 
to be 
the simultaneous eigenstate of $\hat{\mathcal{K}}_{\rm J}$. This can be achieved, if and only if either $\Psi_{j+}=0$ for $j={\rm A},{\rm B},{\rm C}$ or $\Psi_{j-}=0$ for $j={\rm A},{\rm B},{\rm C}$ 
is satisfied. 
Indeed, since Eq.~(\ref{finiteJcondition}) is composed of two equations corresponding to $\alpha=\uparrow$ and $\downarrow$, it is possible to set, at least, three components out of six components 
$\Psi_{j\pm}$ ($j={\rm A},{\rm B},{\rm C}$) to be zero. Consequently, one 
solution of Eq.~(\ref{finiteJcondition}) satisfying $\Psi_{j+}=0$ 
for all $j$ 
is available. This solution belongs to the eigenspace of $\hat{\mathcal{K}}_{\rm J}$ with eigenvalue $J/2$, hence is the eigenstate of $\mathcal{H}$ with eigenvalue $\frac{J}{2}+t$. Similarly, 
another solution with $\Psi_{j-}=0$ 
for all $j$ exists, which has eigenvalue $-\frac{J}{2}+t$ of the Hamiltonian $\mathcal{H}$.
To demonstrate these, we plot the energy spectrum of the Kondo lattice model on a single triangle in Figs.~\ref{TetrahedronLevel}(a) and \ref{TetrahedronLevel}(b) with an arbitrarily chosen spin configuration: ${\mathbf S}_i=(\sin\theta_i\cos\phi_i, \sin\theta_i\sin\phi_i, \cos\theta_i)$,
with $(\theta_{\rm A}, \phi_{\rm A})=(3\pi/5, \pi/10)$, $(\theta_{\rm B}, \phi_{\rm B})=(9\pi/10, 6\pi/5)$ and $(\theta_{\rm C}, \phi_{\rm C})=(2\pi/5, 3\pi/2)$.
As expected, an energy level appears at $\varepsilon=-\frac{J}{2}+t$ and $\frac{J}{2}+t$, respectively.

These results can be generalized to spin-$S$ fermions on a complete graph composed of $n_c$ sites. In this general case, if $n_c-(2S+1)>0$, the energy levels appear at $\varepsilon=JS_z+t$ for $S_z=-S, -S+1, \cdots, S$, with $n_c-(2S+1)$-fold degeneracy for each $S_z$. Otherwise, if $n_c-(2S+1)\leq0$, levels do not exist at these energies. 
As an example, we show the result for $S=3/2$ fermions on a complete graph with $n_c=6$ in [Figs.~\ref{TetrahedronLevel}(c)-\ref{TetrahedronLevel}(f)]. Energy levels appear at $\varepsilon=JS_z+t$ for $S_z=-3/2$, $-1/2$, $1/2$, and $3/2$, each with 
twofold degeneracy, as expected.

\subsection{Extension to a kagome lattice}
\label{kagome}
Now, let us look at how the invariant energy levels manifest themselves, when the complete graphs are connected to form a lattice.
As a representative case, we consider a kagome lattice, which is made by connecting the triangles with sharing their corners [see Fig.~\ref{graph}(b)]. 
We here assume $J=\infty$.

\begin{figure}[h]
\begin{center}
\includegraphics[width=0.9\textwidth]{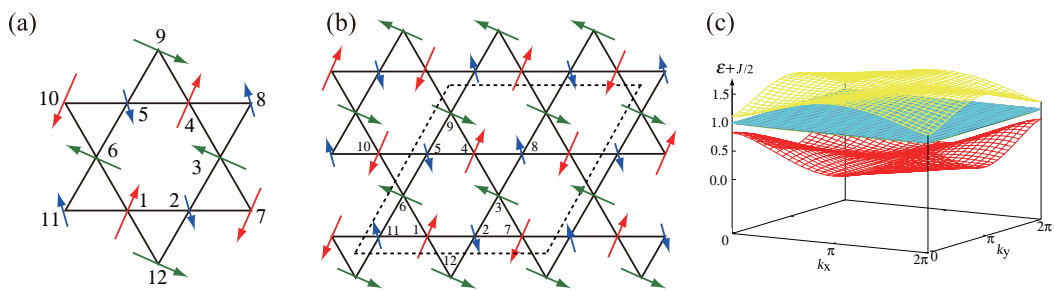}
\end{center}
\caption{\label{config} 
(Color online) (a) An example of spin configuration on a hexagon and surrounding sites, which
leads to a localized mode with eigenenergy $t$. (b) An example of spin configuration on a kagome lattice, which leads to 
a flat band. 
The configuration is constructed by embedding the hexagonal configuration in (b) periodically, with reversing the spin directions properly. 12 sites are contained in a magnetic unit cell shown 
by the dashed line. 
(c) The fifth to tenth lowest-energy bands are shown for the Kondo lattice model (\ref{Ham}) at $J=100$ for 
the ordered spin configuration 
in (b). Spin directions are assumed to be ${\mathbf S}_i=(\sin\theta_i\cos\phi_i, \sin\theta_i\sin\phi_i, \cos\theta_i)$
with $(\theta_1, \phi_1)=(4\pi/5, 2\pi/3)$, $(\theta_2, \phi_2)=(\pi/5, \pi/3)$, and $(\theta_3, \phi_3)=(\pi/5, 0)$. The sixth to ninth bands form almost 
flat bands at $\varepsilon\simeq-\frac{J}{2}+t$ with four-fold degeneracy.
}
\end{figure}

When extending to a lattice, 
the constraint (\ref{constraint}) 
leads to the existence of localized mode around a hexagon of the kagome lattice.
Let us consider a hexagon and surrounding sites on 
the kagome lattice, and assume a spin configuration as shown in 
Fig.~\ref{config}(a). 
Namely, we assume ${\mathbf S}_1$, ${\mathbf S}_2$, and ${\mathbf S}_3$ to be arbitrary, and set ${\mathbf S}_4=-{\mathbf S}_7=-{\mathbf S}_{10}={\mathbf S}_1$, ${\mathbf S}_5=-{\mathbf S}_8=-{\mathbf S}_{11}={\mathbf S}_2$, and ${\mathbf S}_6=-{\mathbf S}_9=-{\mathbf S}_{12}={\mathbf S}_3$.
Note that the surrounding sites $7-12$ have opposite spin directions to the hexagon 
sites $1-6$, respectively.
Given this configuration, 
the one-particle state $|\Psi\rangle = (\Psi_1 a_1^{\dag} + \Psi_2 a_2^{\dag} + \Psi_3 a_3^{\dag} + \Psi_1 a_4^{\dag} + \Psi_2 a_5^{\dag} + \Psi_3 a_6^{\dag})|0\rangle $
becomes the localized eigenstate of the low-energy Hamiltonian (\ref{Heff}), with eigenenergy $\varepsilon=t$, if $\Psi_1$, $\Psi_2$, and $\Psi_3$ satisfy the condition: $\Psi_1\chi_{1\alpha}({\mathbf S}_1) + \Psi_2\chi_{2\alpha}({\mathbf S}_2) + \Psi_3\chi_{3\alpha}({\mathbf S}_3)=0$ for $\alpha=\uparrow$ and $\downarrow$, analogous to 
Eq.~(\ref{constraint}). This 
is easily understood by applying the original Kondo lattice Hamiltonian (\ref{Ham}) to the state $|\Psi\rangle$.
For example, the hopping term $-t\sum_{\alpha}(c_{12\alpha}^{\dag}c_{1\alpha} + c_{12\alpha}^{\dag}c_{2\alpha})$ annihilates electrons at sites $1$ and $2$, and
creates an electron at site $7$ 
in the spin state $\propto -t(\Psi_1|\chi_+({\mathbf S}_1)\rangle + \Psi_2|\chi_+({\mathbf S}_2)\rangle)=t\Psi_3|\chi_+({\mathbf S}_3)\rangle$, which is orthogonal to the low-energy sector at site $12$, since ${\mathbf S}_{12} = -{\mathbf S}_3$ and consequently $|\chi_+({\mathbf S}_{12})\rangle\propto|\chi_-({\mathbf S}_3)\rangle$. 
This destructive interference confines the state within the 
hexagon sites. Moreover, it can be easily shown that the state $|\Psi\rangle$ is, indeed, an eigenstate, by applying the hopping term within the 
hexagon sites. Since this localized state is realized for rather general spin configurations 
[arbitrary sets of (${\mathbf S}_1, {\mathbf S}_2$, 
${\mathbf S}_3$)], it 
may appear in a broad range of spin configurations in a lattice, including the case with random spin configurations and constrained systems as localized or mid-gap states\cite{Ishizuka12,GiaWei12,Udagawa13}.

Among the spin configurations leading to localized modes, a spatially periodic configuration is of special interest, since it gives rise to formation of a flat band, 
analogous to the non-interacting tight-binding model on 
the kagome lattice as discussed in Sec.~\ref{Model}.
Aiming at the engineering of such flat bands, we assume the spin configuration as shown in 
Fig.~\ref{config}(b), 
which is obtained by placing the configuration in 
Fig.~\ref{config}(a) 
in a periodic way, with reversing spin directions properly.
We numerically diagonalized the Hamiltonian (\ref{Ham}), by assuming $J=100$ 
and an arbitrarily chosen spin configuration: ${\mathbf S}_i=(\sin\theta_i\cos\phi_i, \sin\theta_i\sin\phi_i, \cos\theta_i)$
with $(\theta_1, \phi_1)=(4\pi/5, 2\pi/3)$, $(\theta_2, \phi_2)=(\pi/5, \pi/3)$, and $(\theta_3, \phi_3)=(\pi/5, 0)$. In Fig.~\ref{config}(c), we show the obtained energy dispersions.
The figure shows the fifth to tenth lowest-energy bands. The middle ones are almost flat bands 
at $\varepsilon\simeq-\frac{J}{2}+t$, which are isolated from other dispersive bands; 
the flat bands are four-fold degenerate, corresponding to the four hexagons included in the magnetic unit cell [see Fig.~\ref{config}(b)]. 

This construction 
method of flat bands is 
applicable to any triangle-based corner-shared lattices. 
Moreover, since the magnetic order breaks time-reversal symmetry, it is also useful to search for a possible flat band with a nontrivial Chern number. The search for such an exotic flat-band state is left for future study.

\section{Summary} 
To summarize, we have given the detailed derivation of invariant energy levels of the Kondo lattice model on an isolated complete graph. In this model, one-particle energy levels with eigenenergies $\varepsilon=t\pm\frac{J}{2}$ 
appear, irrespective of the configuration of localized moments. This result can be generalized to 
spin-$S$ fermions on a complete graph composed of $n_c$ sites. In this case, the energy levels appear at $\varepsilon=JS_z+t$ for $S_z=-S, -S+1, \cdots, S$, with $n_c-(2S+1)$-fold degeneracy for each $S_z$, provided that $n_c-(2S+1)>0$.
The construction rule 
for these levels gives the existence condition of localized modes in the Kondo lattice model on 
the lattices composed of the corner-sharing complete graphs, such as a kagome lattice. Moreover, we 
establish the way of 
constructing flat bands with broken time-reversal symmetry by connecting the localized modes.
This method will be useful to search for a (nearly-)flat band with nontrivial topological character.

This work was supported by Grant-in-Aid for JSPS Fellows, Grants-in-Aid for Scientific Research (No. 
24340076 
and 24740221), the Strategic Programs for Innovative Research (SPIRE), MEXT, and the Computational Materials Science Initiative (CMSI), Japan.

\end{document}